\documentclass[twocolumn,secnumarabic,amssymb, aps, prl]{revtex4-2}
\usepackage{graphicx}
\usepackage{amsmath}
\usepackage{amsfonts}
\usepackage[hidelinks]{hyperref}

\makeatletter
\def\maketitle{
\@author@finish
\title@column\titleblock@produce
\suppressfloats[t]}
\makeatother

\begin{document}

\title{Emergent Quasiperiodicity from Polariton-phonon Hybrid Excitations in Waveguide Quantum Optomechanics}%

\author{Han-Jie Zhu$^{1}$}
\author{Xiao-Ming Zhao$^{2}$}
\author{Jin-Kui Zhao$^{1,6}$}
\author{Lin Zhuang$^{3}$}
\email{stszhl@mail.sysu.edu.cn}
\author{Guo-Feng Zhang$^{4}$}
\email{gf1978zhang@buaa.edu.cn}
\author{Wu-Ming Liu$^{1,5,6}$}
\email{wliu@iphy.ac.cn}
\affiliation{$^{1}$Beijing National Laboratory for Condensed Matter Physics, Institute of Physics, Chinese Academy of Sciences, Beijing 100190, China}
\affiliation{$^{2}$Department of Physics and Institute of Theoretical Physics, University of Science and Technology Beijing, Beijing 100083, China}
\affiliation{$^{3}$State Key Laboratory of Optoelectronic Materials and Technologies, School of Physics, Sun Yat-Sen University, Guangzhou 510275, China}
\affiliation{$^{4}$School of Physics, Beihang University, Beijing 100191, China}
\affiliation{$^{5}$School of Physical Sciences, University of Chinese Academy of Sciences, Beijing 100190, China}
\affiliation{$^{6}$Songshan Lake Materials Laboratory, Dongguan, Guangdong 523808, China}

\begin{abstract}
We investigate polariton-phonon hybrid excitations, 
which describe the collective excitations of emitter-photon polaritons and vibrational phonons, 
in a periodic array of vibrating two-level emitters interacting with waveguide photons. 
We demonstrate the emergence of an interaction-induced quasiperiodic structure caused by the interplay between phonon scatterings and waveguide-mediated long-range couplings. 
This quasiperiodicity fundamentally changes the excitation characteristics in the subradiant regime, which feature an appearance of topological edge states and a transition between ergodic and multifractal excitations. 
A possible realization consisting of an array of laser-cooled atoms trapped near an optical nanofiber is also proposed. 
Our results demonstrate the possibility of utilizing vibrations as a novel degree of freedom in the exploration of many-body physics with waveguide quantum electrodynamics systems.
\end{abstract}

\maketitle

Waveguide quantum electrodynamics (QED), an emerging field focusing on the interaction of propagating waveguide photons with quantum emitters, has attracted intense interest in recent years motivated by the significant progresses in quantum technologies \cite{Optical_nanofibres_and_neutral_atoms,Strongly_interacting_photons_in,
Quantum_matter_built_from_nanoscopic_lattices,Coherent_nonlinear_optics_of_quantum,Waveguide_quantum_electrodynamics_collective,
Corzo2019,PhysRevLett.126.023603,An_integrated_diamond_nanophotonics,Mirhosseini2019}. 
Besides applications in quantum network and quantum computation, waveguide QED also provides a promising platform for quantum simulations of many-body physics. 
Such systems present an unusual behavior due to strong light-matter interactions and long-range couplings between emitters mediated by waveguide photons \cite{Exponential_Improvement_in_Photon_Storage,Albrecht2019,PhysRevA103033702,PhysRevA104013303}. 
These features allow the exploration of various exotic phenomena, including 
unconventional topological phases \cite{Quantum_Electrodynamics_in_a_Topological_Waveguide,PhysRevLett124083603}, 
superradiant and subradiant states \cite{PhysRevA101043845,PhysRevResearch2043149,
PhysRevLett123253601,PhysRevLett124213601,PhysRevLett125253601,PhysRevLett126223602,PhysRevLett127263602,PhysRevX11021031,
PhysRevLett128113601,PhysRevLett128093602,WOS000788592600025,WOS000768639500001,WOS000601155900006,WOS000503748600005,WOS000731422500004,WOS000704666000001,WOS000517416700006,PhysRevA106L031702}, 
and peculiar correlations between photons \cite{PhysRevA93033856,Prasad2020,PhysRevLett126083605,PhysRevLett127273602}. 

While emitters are assumed to be static in most waveguide QED studies, their mechanical motions can act as novel degrees of freedom due to the position-dependent nature of light-matter interactions \cite{PhysRevLett110113606,PhysRevA102013726}. 
In particular, the vibrational degrees of freedom, which arise naturally in cold-atom experiments \cite{Bose_Einstein_condensation}, 
can contribute to interesting phenomena even in small systems with few emitters 
\cite{PhysRevLett125183601,PhysRevLett125263606,PhysRevA104063719}. 
In the many-body regime (many emitters), the collective excitations of emitters and photons (polaritons) follow unconventional dispersion relations 
and exhibit peculiar effects ranging from fermionization to quantum chaos \cite{PhysRevLett122203605,PhysRevLett124093604,Poshakinskiy2021,PhysRevResearch2013173,PhysRevLett127173601,PhysRevLett126203602,WOS000548060300001,PhysRevA103023720}. 
These polaritons can propagate along the waveguide and interact with lattice vibrations, thus the resulting hybrid excitations may behave quite differently compared to the bare polaritons. 
We can expect the emergence of highly interesting many-body phenomena as a result of the interplay between waveguide polaritons and vibrational phonons.
Nevertheless, the physics resulting from polariton-phonon interactions remains largely unexplored in waveguide QED due to the complexity from the hybridization of photons, emitters, and phonons, as well as the long-range nature of the photon-mediated interactions. 

In this Letter, we develop a description of polariton-phonon hybrid excitations in an array of vibrating emitters coupled to a waveguide. 
We find that these excitations exhibit a distinctly different behavior compared to bare polaritons in a subwavelength finite array. 
We then identify the emergence of an effective quasiperiodic potential, which is caused by the combination of phonon scatterings and waveguide-mediated long-range couplings, as the key element behind the intriguing behavior of hybrid excitations. 
This quasiperiodicity splits the originally continuous spectrum into a set of bands, and gives rise to topological edge states inside the bandgaps. 
In addition, the subradiant excitations experience an ergodic-multifractal transition with an edge 
separating the spectrum into two regions due to the quasiperiodicity. 
We also suggest a cold-atom realization to explore our predicted phenomena. 

\begin{figure}
            \centering
            \includegraphics[width=\columnwidth]{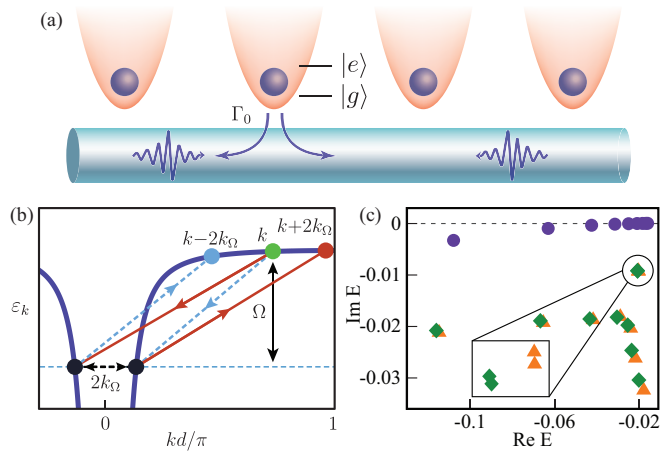}
            \caption{
               (a) Schematic illustration of a periodic array of vibrating two-level emitters trapped near a waveguide and interacted with propagating photons. (b) Lower branch of the single-polariton dispersion. Phonons can mediate effective interactions between polaritons with quasi-momentum change $\pm 2 k_{\Omega}$. (c) Subradiant part of the complex single-excitation spectrum obtained from the Schrieffer-Wolff transformation (orange triangle) in comparison with the exact diagonalization results (green diamond) and the single-polariton spectrum (blue circle). The inset shows the detailed structure of the near-degenerate group in spectrum. The calculation has been performed for an array of $N=12$ emitters with $\Gamma_{0}=\Omega=1$ and $\eta=\varphi=\pi / 50$.
            }
            \label{Fig1}
\end{figure}

We consider a periodic array of $N$ traps along a one-dimensional waveguide, each loaded with a two-level emitter, as depicted in Fig. \ref{Fig1}(a). The emitters strongly radiate into the waveguide and are allowed to vibrate parallel to the waveguide. The absorption or emission of waveguide photons by emitters can lead to the deformation of the array structure, thus excite the vibrational modes (phonons). 
The system is characterized by \cite{PhysRevLett125183601}
\begin{equation}
\begin{aligned}
\mathcal{H}=& \sum_{k} \omega_{k} b_{k}^{\dagger} b_{k}+\sum_{m} \omega_{0} \sigma_{m}^{\dagger} \sigma_{m}+\sum_{m} \Omega a_{m}^{\dagger} a_{m} \\
&+\frac{g}{\sqrt{L}} \sum_{k, m}\left(\sigma_{m}^{\dagger} b_{k} e^{i k \hat{z}_{m}}+\sigma_{m} b_{k}^{\dagger} e^{-i k \hat{z}_{m}}\right),
\end{aligned}
\end{equation}
where $b_{k}, \sigma_{m}$, and $a_{m}$ are the annihilation operators of the waveguide photon, emitter excitation, and phonon of the $m$-th site respectively. Here, $\omega_{k}=c|k|$ is the frequency of the photon with wave vector $k$, where $c$ is the light speed in the waveguide. The resonance frequencies of the emitters and vibrational modes are given by $\omega_{0}$ and $\Omega$ respectively. Parameter $g$ is the atom-light interaction strength, $L$ is the normalization length, and $\hat{z}_{m}=z_{m}+u_{0} \hat{x}_{m}$ is the position operator where $z_{m}$ is the equilibrium position of the $m$-th atom, $u_{0}$ is the quantum of the vibrational mode, and $\hat{x}_{m}=a_{m}^{\dagger}+a_{m}$. 
Instead of few-emitter cases, we focus on the many-body regime ($N \gg 1$) where emitters and photons form polaritons 
with strong collective superradiant and subradiant behavior. 

In the Markovian approximation, photons can be integrated out \cite{Waveguide_quantum_electrodynamics_collective,PhysRevA84063803}
and the system is described by the effective Hamiltonian $H=H_{0}+H_{p}+H_{I}$, where $H_{0}=-i \Gamma_{0} \sum_{m, n} e^{i \varphi|m-n|} \sigma_{m}^{\dagger} \sigma_{n} / 2$ and $H_{p}=\sum_{m} \Omega a_{m}^{\dagger} a_{m}$ are the Hamiltonians of emitters and phonons respectively, and
\begin{equation}
H_{I}=-i \frac{\Gamma_{0}}{2} \sum_{m, n} e^{i \varphi|m-n|+i \eta \operatorname{sign}(m-n)\left(\widehat{x}_{m}-\widehat{x}_{n}\right)} \sigma_{m}^{\dagger} \sigma_{n}-H_{0},
\end{equation}
where $\Gamma_{0}=2 g^{2} / c$ is the decay rate for a single emitter into the waveguide, the phase $\varphi \equiv k_{0} d \, ( \bmod 2 \pi )$ is determined by the wave number $k_{0}=\omega_{0} / c$ and the spacing $d$ between adjacent emitters, and $\eta=k_{0} u_{0}$ is the relative optomechanical coupling. The emitter excitation number $\mathcal{N}=\sum_{m} \sigma_{m}^{\dagger} \sigma_{m}$ is conserved in $H$, thus the Hamiltonian can be projected to subspaces with fixed $\mathcal{N}$, where the free Hamiltonian of atoms $\sum_{m} \omega_{0} \sigma_{m}^{\dagger} \sigma_{m}$ produces constant energy and can be discarded. We restrict ourselves to the Lamb-Dicke regime $\eta \langle\widehat{x}_{m}^{2}\rangle^{1 / 2} \ll 1$, where the mechanical fluctuations are small enough compared to the atomic spacing, i.e., $u_{0} \langle\widehat{x}_{m}^{2}\rangle^{1 / 2} \ll d$. In this regime, the single-phonon process is dominant, thus we can neglect processes where multiple phonons are simultaneously absorbed or emitted 
\cite{PhysRevLett70556,Pedernales2015,PhysRevLett118073001,PhysRevLett125050402}.

We first consider an infinite array $(N \rightarrow \infty)$ which is invariant under lattice translations. 
Without phonons, the single-excitation eigenstates are light-matter excitations (polaritons) $|k\rangle=\sigma_{k}^{\dagger}|0\rangle=$ $N^{-1/ 2} \sum_{m} e^{i k z_{m}} \sigma_{m}^{\dagger}|0\rangle$ with quasi-momentum $k$ \cite{Exponential_Improvement_in_Photon_Storage,PhysRevLett122203605}. 
The states $\left|\pm k_{0}\right\rangle$ emit photons superradiantly to the waveguide and are marked by a large imaginary eigenvalue $-i N \Gamma_{0} / 4$, while the remaining $N-2$ states are dark with zero decay rate. The energy dispersion of dark states is given by $\varepsilon_{k}=\left(\Gamma_{0} / 4\right.
) \sum_{\epsilon=\pm} \cot \left[\left(k_{0}+\epsilon k\right) d / 2\right]$. Its curve is split into upper and lower branches separated by a gap. 

In the presence of atom-phonon interactions, polaritons and phonons are coupled to each other and form hybrid excitations. These excitations $\left(k \neq \pm k_{0}\right)$ acquire zero decay rates and remain non-radiative at $O\left(\eta^{2}\right)$ level 
\cite{Supplimentary_material}.
Meanwhile, we focus on the subradiant excitations, which locate near $k=\pm \pi / d$ and present a quadratic dispersion relation. 
In the subwavelength limit $\varphi \ll 1$, the effective mass can be obtained as $m^{*} \approx\left(1+ \eta^{2} \Gamma_{0} /(\varphi \Omega) \right)^{-1} m$, where the bare polariton mass $m$ is 
determined by $\varepsilon_{k}-\varepsilon_{\pi} \approx-\varphi \Gamma_{0}(k d-\pi)^{2} / 16$ near $k=\pi / d$. 
Therefore, the excitation becomes lighter due to the phonon dressing.

While the dressing effects of phonons are relatively simple in infinite cases, they can lead to non-trivial phenomena in finite arrays. 
In Fig. \ref{Fig1}(c), we plot the energy spectrum of the lower excitation branch obtained by exact diagonalization. 
We find that the spectrum of hybrid excitations deviates significantly from the bare polaritons. Its subradiant part shows irregular characters, and a group of near-degenerate states separated from the spectrum branch can be identified.

To investigate phonon effects in longer arrays, we decouple polaritons from phonons via the Schrieffer-Wolff transformation (SWT) \cite{BRAVYI20112793,coleman_2015}. 
An effective Hamiltonian $H^{\prime}=P_{0} e^{S} H e^{-S} P_{0}$ can be obtained by choosing a proper operator $S$ to eliminate the polariton-phonon coupling to the first order, 
where $P_{0}$ is the projector onto the single-polariton subspace without phonons \cite{Supplimentary_material}. 
In finite arrays, dark states become subradiant and possess complex eigenvalues. 
The discrepancy in their decay rates creates an imaginary energy gap separating the single-polariton subspace without phonons from the rest of the spectrum.
This feature allows the decoupling of emitters from phonons as long as the coupling is much smaller than the energy gap.
In the limit of large $N$, this is equivalent to $\eta^2 \varphi^{-1}\left(\Gamma_0 / \Omega\right)^3 \ll N^{-1}$.
This condition can always be satisfied in the weak coupling regime $\eta \ll 1$ by adjusting the decay rate $\Gamma_0$ and the atomic spacing $d$, which are highly tunable in experiments.
More details about the validity of SWT can be found in the Supplemental Material \cite{Supplimentary_material}. 
In Fig. \ref{Fig1}(c), we demonstrate the accuracy of SWT by comparing the energy spectra obtained from $H^{\prime}$ with the exact diagonalization results. 
Here a satisfactory agreement is obtained between the two methods. 

\begin{figure}
            \centering
            \includegraphics[width=\columnwidth]{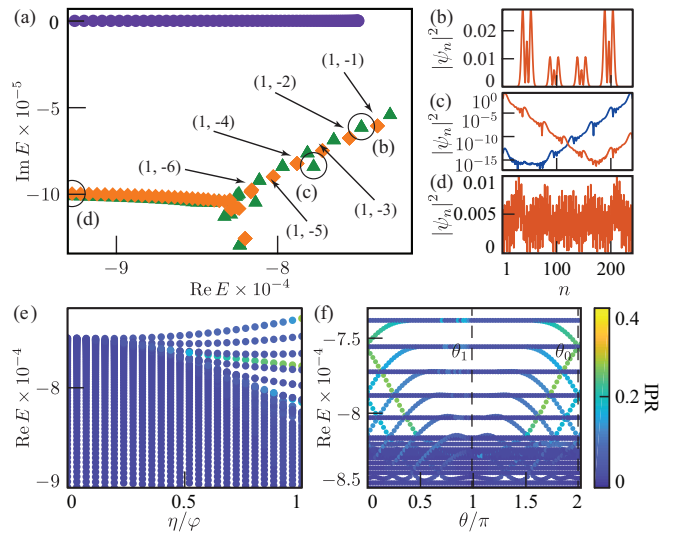}
            \caption{
               (a) The most subradiant part of the complex single-excitation spectrum obtained from the effective Hamiltonian Eq. (\ref{effective_Hamiltonian}) (orange diamond) in comparison with the Schrieffer-Wolff transformation results (green triangle) and the single-polariton spectrum (blue circle). Topological invariants $(\mu, \nu)$ for clearly observable spectral gaps are indicated. (b, c, d) The wavefunctions of three characteristic eigenstates corresponded to the states (b), (c), and (d) in spectrum (a) respectively. (e) Real spectrum of the most subradiant excitations as a function of the optomechanical coupling $\eta$. (f) Real spectrum of the most subradiant excitations in $H_{\text {eff }}$ (Eq. (\ref{effective_Hamiltonian})) with the same parameters in (a), as a function of the modulation phase $\theta$. 
In real arrays, $\theta$ is fixed at discrete values $\theta_{n} = n \pi-\pi \beta(N+1) \bmod 2 \pi$ due to the mirror symmetry inherited from the original Hamiltonian $\mathcal{H}$. 
The calculation has been performed for an array of $N=240$ emitters with $10\Gamma_{0}=\Omega=1, \varphi=0.03$ and $\eta / \varphi=1$ (except in (e)).
            }
            \label{Fig2}
\end{figure}

In Figs. \ref{Fig2}(a) and (e), we present the excitation spectrum in a longer array, where its most subradiant part splits into narrow bands with the increase of $\eta$, 
as opposed to the original polariton spectrum which is continuous and composed of delocalized Bloch states. 
For each normalized eigenstate $|\psi\rangle=\sum_{n} \psi_{n}|n\rangle$, we calculate its inverse participation ratio (IPR) as a measure of localization, which is defined as IPR $=\sum_{n}\left|\psi_{n}\right|^{4}$ and crosses from $N^{-1}$ in the completely delocalized regime to 1 in the localized situation.
The spectrum consists of different types of excitations with diverse spatial distribution, as shown in Figs. \ref{Fig2}(b-d). 
Compared to the delocalized states in the continuous band (Fig. \ref{Fig2}(d)), states in the narrow bands become spatially modulated (Fig. \ref{Fig2}(b)). Moreover, we identify degenerate pairs of edge states which are highly localized at boundaries (Fig. \ref{Fig2}(c)). 
These features indicate that the subradiant excitations are significantly modified by phonons in the finite arrays. 

Our central objective is to identify the mechanism responsible for these intriguing behaviors. 
Therefore, we consider a finite array in the $\varphi \ll 1$ limit, and focus on the subradiant states on the lower excitation branch. 
The subradiant states in infinite arrays with energy $E_{k}$ can be approximated by $\left|\Psi_{k}\right\rangle=|k, 0\rangle+$ $\sum_{p} A_{k}(p)|k-p, p\rangle$, where $|k, 0\rangle=\sigma_{k}^{\dagger}|0\rangle$ and $|k, p\rangle=\sigma_{k}^{\dagger} a_{p}^{\dagger}|0\rangle$ are Bloch states without and with phonons respectively \cite{boundary_effect}. 
$A_{k}(p)$ has two peaks at $p=k \pm k_{\Omega}$, which correspond to the resonant phonon processes connecting $|k, 0\rangle$ to $\left|\mp k_{\Omega}, k \pm k_{\Omega}\right\rangle$ respectively, where $k_{\Omega}>0$ is determined by $\varepsilon_{k}=\Omega+\varepsilon_{k_{\Omega}}$. 
By acting $H$ on this state, we obtain the eigenvalue relation $H\left|\Psi_{k}\right\rangle=E_{k}\left|\Psi_{k}\right\rangle-i \Gamma_{0} \Delta_{k} / 2$ where $\Delta_{k}$ is the boundary term
\begin{equation}
\Delta_{k}=\Delta_{k, 0}+\sum\nolimits_{\pm} A_{k}\left(k \pm k_{\Omega}\right) \Delta_{\mp k_{\Omega}, k \pm k_{\Omega}}.
\end{equation}
Here $\Delta_{k, 0}=g_{k}\left|k_{0}, 0\right\rangle-h_{k}\left|-k_{0}, 0\right\rangle$ and $\Delta_{\pm k_{\Omega}, k \mp k_{\Omega}}=g_{\pm k_{\Omega}}\left|k_{0}, k \mp k_{\Omega}\right\rangle-h_{\pm k_{\Omega}}\left|-k_{0}, k \mp k_{\Omega}\right\rangle$ describe the boundary effects on $|k, 0\rangle$ and $\left|\pm k_{\Omega}, k \mp k_{\Omega}\right\rangle$ respectively, and the coefficients are $g_{k}=$ $e^{i\left(k-k_{0}\right) z_{1}} /\left[1-e^{i\left(k-k_{0}\right) d}\right]$ and $h_{k}=e^{i\left(k+k_{0}\right) z_{N}} /\left[e^{-i\left(k+k_{0}\right) d}-1\right]$. 
The boundary term $\Delta_{k}$ can be canceled by the linear combination of $|k, 0\rangle$ and $|-k, 0\rangle$, and this procedure provides the correct eigenstate $g_{-k}|k, 0\rangle-g_{k}|-k, 0\rangle$ in the absence of atom-phonon coupling, where the wavenumber is given by the equation $g_{k} h_{-k}=g_{-k} h_{k}$ \cite{PhysRevLett122203605}. 
To cancel $\Delta_{\pm k_{\Omega}, k \mp k_{\Omega}}$, we notice that $k_{\Omega}$ behaves almost as a constant and can be approximated by $k_{\Omega} d \approx\left(\varphi \Gamma_{0} / \Omega\right)^{1 / 2}$, 
as long as $|k\rangle$ remains in the quasi-flat regime on the lower excitation branch, as shown in Fig. \ref{Fig1}(b). 
This effectively induces a coupling between $\left|\Psi_{k}\right\rangle$ and $\left|\Psi_{k \pm 2 k_{\Omega}}\right\rangle$.

Physically, this effective coupling can be understood as follows: 
Polaritons can be reflected from the boundaries in finite arrays. 
A polariton $|k, 0\rangle$ can be scattered into its resonant state $\left|\pm k_{\Omega}, k \mp k_{\Omega}\right\rangle$ and emits a phonon, while this state can be reflected into $\left|\mp k_{\Omega}, k \mp k_{\Omega}\right\rangle$, and finally scattered back into $\left|k \mp 2 k_{\Omega}, 0\right\rangle$ after absorbing the phonon previously emitted, as shown in Fig. \ref{Fig1}(b). This process effectively creates an interaction with momentum change $\mp 2 k_{\Omega}$.

This momentum-change process plays a crucial role in the intriguing behavior of subradiant excitations. 
In the $\varphi \ll 1$ regime, we obtain an approximated expression of $H^{\prime}$ as $H^{\prime} \approx H_{0}+\Delta H$, where the phonon-induced interactions are $\Delta H=\sum_{k}\left[V_{k}\left(e^{-i \theta_{k}}|k\rangle\left\langle k+2 k_{\Omega}\right|+\right.\right.$ $\left.\left.e^{i \theta_{k}}\left|k+2 k_{\Omega}\right\rangle\langle k|\right) / 2+\delta \varepsilon_{k}|k\rangle\langle k|\right]$. Here $V_{k}$ is a complex effective coupling strength, $\theta_{k}$ is a phase factor, and $\delta \varepsilon_{k}$ accounts for the energy shift \cite{Supplimentary_material}. 
All three coefficients change slowly in the quasi-flat regime, 
thus $H^{\prime}$ can be further approximated by replacing three coefficients by their values at $k=\pi / d$.
In real space, the resulting effective Hamiltonian can be written as
\begin{equation}
H^{\prime} \approx H_{e f f}=H_{0}+V \sum_{m} \cos \left(2 k_{\Omega} d m+\theta\right)|m\rangle\langle m|,
\label{effective_Hamiltonian}
\end{equation}
which describes the original polaritons with an additional complex on-site potential being cosine modulated. Here $V$ and $\theta$ serve as the amplitude and phase of the modulation, and take the values of $V_{k}$ and $\theta_{k}$ at $k=\pi / d$ respectively \cite{Supplimentary_material}. 
Meanwhile, we omit the near-constant energy shift $\varepsilon_{k}$. 
This effective Hamiltonian provides a qualitatively correct description of subradiant excitations, as shown in Fig. \ref{Fig2}(a).

The on-site potential in $H_{eff}$ (Eq. (4)) is in general incommensurate with the lattice since $k_{\Omega}$ is determined by the transcendental equation $\varepsilon_{k}=\Omega+\varepsilon_{k_{\Omega}}$. 
Therefore, $H_{eff}$ can be regarded as a one-dimensional quasiperiodic model with long-range waveguide-mediated hoppings and a complex potential. 
This quasiperiodicity fundamentally changes the behavior of subradiant excitations. 
We notice that $k_{\Omega}$ can be approximated by $k_{\Omega} d / \pi \approx q^{-1}$ where $q \gg 1$ is an integer since $k_{\Omega} d / \pi \ll 1$. 
Thus, the quasiperiodicity naturally splits the spectrum into $q$ bands in the first-order rational approximation, and the bands near the edge of the spectrum become much narrower. This explains the splitting of the spectrum in Figs. \ref{Fig2}(a) and (e). 

Interestingly, the spectrum includes degenerate pairs of edge states localized over boundaries (Fig. \ref{Fig2}(c)). 
These states are topological edge states as a result of the quasiperiodicity. 
This is revealed through $H_{e f f}$, which inherits the topological properties of its two-dimensional ancestor Hamiltonian $H_{2 D}=\int_{0}^{2 \pi}(d \theta / 2 \pi) H_{e f f}(\theta)$. 
Here $\theta$ is regarded as a momentum in a perpendicular synthetic dimension, and $\sigma_{m}$ is replaced by $\sigma_{m, \theta}$ in $H_{e f f}(\theta)$. 
By performing the Fourier transform $\sigma_{m, \theta}=\sum_{l} e^{-i \theta l} \sigma_{m, l}$, we have 
$H_{2 D}=-i\left(\Gamma_{0} / 2\right) \sum_{m n l} e^{i \varphi|m-n|} \sigma_{m, l}^{\dagger} \sigma_{n, l}+(V / 2) \sum_{m l}(e^{i 2 k_{\Omega} d m} \sigma_{m, l}^{\dagger} \sigma_{m, l+1}+h.c.)$. 
Similar to conventional quasicrystals \cite{PhysRevLett109116404}, the ancestor Hamiltonian $H_{2 D}$ commutes with the magnetic translation group generated by $T_{m}$ and $T_{l}$, where $T_{l} \sigma_{m, l} T_{l}^{-1}=\sigma_{m, l+1}$ and $T_{m} \sigma_{m, l} T_{m}^{-1}=$ $e^{-i 2 k_{\Omega} d m} \sigma_{m+1, l}$. 
Thus, each gap in the spectrum of $H_{2 D}$ can be characterized by a quantized and nontrivial Chern number, and this feature is inherited by $H_{e f f}$. 
As a result, the band topology of $H_{e f f}$ can be described by Chern numbers, which satisfies the Diophantine equation $\rho=\mu+\nu \left(k_{\Omega} d / \pi\right)$ where $\mu$ is an integer, $\nu$ is the Chern number, and $\rho$ is the filling factor within a gap \cite{PhysRevA103013727,Zilberberg21,WOS000383955500004,WOS000553582900001}. 
For an irrational $k_{\Omega} d / \pi$, the Diophantine equation has only one solution when $\rho$ is fixed, thus each gap can be labeled by a set of integers $(\mu, \nu)$. 

Taking Fig. \ref{Fig2}(a) as an example, we find that the difference between Chern numbers of neighboring gaps is 1, thus each gap produces a pair of edge states localized on two edges respectively according to the bulk-boundary correspondence \cite{PhysRevLett109106402}. 
This agrees with the number of edge states in Fig. \ref{Fig2}(f), and confirms the topological nature of edge states. 
Furthermore, the modulation phase $\theta$ in real arrays can only take discrete values $\theta_{n}=n \pi-\pi \beta(N+1)$ due to the mirror symmetry inherited from the original Hamiltonian $\mathcal{H}$. 
As a result, edge states on opposite edges always form degenerate pairs, which is consistent with Fig. \ref{Fig2}(a).

Besides edge states, the quasiperiodicity also changes the ergodic nature of subradiant excitations, 
which is revealed through the analysis of the even-odd (odd-even) energy spacing $S_{n}^{e-o}=E_{2 n}-E_{2 n-1}$ ($S_{n}^{o-e}=E_{2 n+1}-E_{2 n}$), 
where $E_{n}$ are the real eigenenergy parts sorted in ascending order \cite{PhysRevLett123025301,PhysRevB103184309}. 
For multifractal states, the spacings exhibits a strongly scattering pattern, while ergodic states possess regular and continuous spacings \cite{explanation_energy_spacing}. 
When $\eta$ is small, all excitations are ergodic (Fig. \ref{multifractal}(a)). 
For large $\eta$ (Fig. \ref{multifractal}(b)), bands in the most subradiant regime become multifractal, and the spectrum presents an edge separating the ergodic and multifractal excitations \cite{Supplimentary_material}.
Such a transition is the result of the interplay between long-ranged hoppings and the effective quasiperiodic potentials. 
This is confirmed via multifractal analysis on $H_{e f f}$ that shows it exhibits similar multifractal behavior to our system \cite{Supplimentary_material}. 

\begin{figure}
            \centering
            \includegraphics[width=\columnwidth]{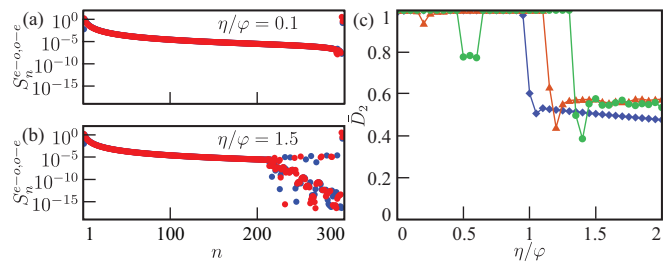}
            \caption{
               (a, b). Level spacing $S_{n}^{e-o}$ (red) and $S_{n}^{o-e}$ (blue) for the system with $\eta / \varphi=0.1$ (a) and $\eta / \varphi=1.5$ (b) respectively. (c). Mean fractal dimensions $\bar{D}_{2}$ for three subbands with the highest energy. 
Three highest subbands arranged in energy-descending order are labeled by blue diamond, red triangle and green circle respectively. All of these figures were generated for an array of $N=600$ emitters with $10\Gamma_{0}=\Omega=1$ and $\varphi=0.03$.
            }
            \label{multifractal}
\end{figure}

The ergodic-multifractal transition is further confirmed by the analysis of fractal dimensions \cite{PhysRevB68024206,PhysRevLett104070601,RevModPhys93045001,PhysRevB104224204,PhysRevB100144202,WOS000505617400023,WOS000817765700006}. A normalized wavefunction $|\psi\rangle=\sum_{n} \psi_{n}|n\rangle$ can be characterized by the moments $I_{q}=$ $\sum_{n}\left|\psi_{n}\right|^{2 q} \propto N^{-D_{q}(q-1)}$ where $D_{q}$ are fractal dimensions. For ergodic (localized) excitations, $D_{q}=1$ $\left(D_{q}=0\right)$, while $D_{q} \in(0,1)$ for multifractal excitations. In Fig. \ref{multifractal}(c), we show the mean fractal dimensions $\bar{D}_{2}$ over excitations within the same bands. For each band, there exists a critical coupling $\eta$ at which excitations within this band exhibit multifractal behavior. Thus, subradiant bands become multifractal in sequence when $\eta$ is increased.

For experimental realizations, we consider an array of laser-cooled atoms trapped near an optical nanofiber.
In this setup, the phonon frequency $\Omega$ is around several $\mathrm{MHz}$, while the decay rate $\Gamma_{0}$ is highly tunable and can be adjusted to the same order of magnitude as $\Omega$ \cite{Gerritsma2010,WOS000492373300003,WOS000417479200001,Corzo2019,WOS000355281400006,WOS000384108600003}.
A typical value of the optomechanical coupling is $\eta \sim 0.05$ for cesium atoms with transition energy $\hbar \omega_{0} \sim 1.4 \mathrm{eV}$ at $\Omega \sim 1 \mathrm{MHz}$. 
Therefore, the parameter ranges that we consider are accessible in experiments. 
In the subwavelength limit ($k_{0} d \ll 1$), the restriction $\eta \ll k_{0} d=\varphi$ should be imposed since the vibrations around the equilibrium positions are assumed to be much smaller than the atomic spacing. Nevertheless, the energy spectrum is already significantly modified in this regime and exhibits visible gaps manifesting the quasiperiodicity.
To explore the $\eta \sim \varphi$ regime, a near-Bragg-spaced atomic array with $k_{0} d=2 \pi+\varphi$ may be applied.

In conclusion, we have shown the emergence of polariton-phonon hybrid excitations with novel properties in a waveguide QED system due to the interplay between waveguide-mediated long-range couplings and phonon scatterings. 
In a subwavelength finite array, these interactions introduce an effective quasiperiodic structure which leads to the appearance of topological edge states and a transition between ergodic and multifractal excitations. 
The emergent quasiperiodicity provides an interesting playground for studying one-dimensional quasicrystals due to the infinite-ranged nature of waveguide-mediated couplings, which are vastly different from the tight-binding or long-ranged couplings in typical quasicrystals. 
Our work sheds light on the mechanism of interaction between waveguide polaritons and vibrational phonons, 
and motivates future investigations into exploring many-body physics with waveguide QED systems utilizing emitter vibrations. 
It also provides insight into the understanding of polariton-phonon
interactions in other fields where hybridized polaritons and phonons are important, e.g., in atomic Bose-Einstein condensates \cite{PhysRevLett116053602,PhysRevLett125035301} and polariton chemistry \cite{C8SC01043A}.

\begin{acknowledgments}
This work was supported by National Key R\&D Program of China under grants No. 2021YFA1400900, 2021YFA0718300, 2021YFA1400243, 
NSFC under grants Nos. 12074027, 61835013, 12174461, 12234012, Space Application System of China Manned Space Program, 
the Recruitment Program of Global Experts, 
and China Postdoctoral Science Foundation under grants No. 2020M680725.
\end{acknowledgments}

        \bibliography{paper.bib}

\begin{thebibliography}{84}
\bibitem [{bou()}]{boundary_effect}%
  \BibitemOpen
  \href@noop {} {}\bibinfo {note} {Here the wavefunction is truncated at
  single-phonon level since single-phonon processes play a primary role within
  the Lamb-Dicke regime. Meanwhile, in the eigenvalue relation we accounts for
  the boundary effects of $H_{0}$ only since the emitter-phonon couplings
  create much smaller boundary effects. Similarly, the contributions from the
  single-phonon states are also neglected except for $\left|\pm k_{\Omega}, k
  \mp k_{\Omega}\right\rangle$}\BibitemShut {NoStop}%
\bibitem [{exp()}]{explanation_energy_spacing}%
  \BibitemOpen
  \href@noop {} {}\bibinfo {note} {With periodic boundary conditions, ergodic
  states are doubly degenerate and a gap is expected between $S_{n}^{e-o}$ and
  $S_{n}^{o-e}$ in ergodic regions. This degeneracy is lifted with open
  boundaries, thus the gap vanishes in Figs. 3(a) and (b).}\BibitemShut {Stop}%
\end{thebibliography}

\clearpage


    \makeatletter
    \renewcommand{\theequation}{S\arabic{equation}}
\setcounter{equation}{0}
    \renewcommand{\thefigure}{S\arabic{figure}}
\setcounter{figure}{0}
    \renewcommand{\thesection}{S\arabic{section}}


\onecolumngrid

\section{Calculation of infinite arrays}
In this section, we present a detailed derivation for the results in infinite arrays. We start from the effective Hamiltonian $H$ in the main text
\begin{equation}
H=\sum_{m} \Omega a_{m}^{\dagger} a_{m}-i \frac{\Gamma_{0}}{2} \sum_{m, n} e^{i \varphi|m-n|+i \eta \operatorname{sign}(m-n)\left(a_{m}+a_{m}^{\dagger}-a_{n}-a_{n}^{\dagger}\right)} \sigma_{m}^{\dagger} \sigma_{n}.
\end{equation}
In the Lamb-Dicke regime $\eta \sqrt{\left\langle\left(a_{m}+a_{m}^{\dagger}\right)^{2}\right\rangle} \ll 1$, the exponential term can be expanded as
\begin{equation}
\begin{gathered}
e^{i \eta \operatorname{sign}(m-n)\left(a_{m}+a_{m}^{\dagger}-a_{n}-a_{n}^{\dagger}\right)} \approx 1+i \eta \operatorname{sign}(m-n)\left(a_{m}+a_{m}^{\dagger}-a_{n}-a_{n}^{\dagger}\right) \\
-\frac{\eta^{2}}{2}\left(a_{m}+a_{m}^{\dagger}-a_{n}-a_{n}^{\dagger}\right)^{2}+O\left(\eta^{3}\left(a_{m}+a_{m}^{\dagger}-a_{n}-a_{n}^{\dagger}\right)^{3}\right).
\end{gathered}
\end{equation}
Therefore, the Hamiltonian $H$ can be approximated up to second order in the emitter displacements as
\begin{equation}
\begin{aligned}
H \approx \sum_{m} \Omega a_{m}^{\dagger} a_{m} &-i \frac{\Gamma_{0}}{2} \sum_{m, n} e^{i \varphi|m-n|} \sigma_{m}^{\dagger} \sigma_{n}\left[1+i \eta \operatorname{sign}(m-n)\left(a_{m}+a_{m}^{\dagger}-a_{n}-a_{n}^{\dagger}\right)\right.\\
&\left.-\frac{1}{2} \eta^{2}\left(a_{m}+a_{m}^{\dagger}-a_{n}-a_{n}^{\dagger}\right)^{2}\right].
\end{aligned}
\end{equation}
In the infinite array limit $N \rightarrow \infty$, we can transform the Hamiltonian into the $k$-space by defining the Fourier transform of the operators as
\begin{equation}
\sigma_{k}=\frac{1}{\sqrt{N}} \sum_{m} e^{-i k m d} \sigma_{m}, \quad a_{p}=\frac{1}{\sqrt{N}} \sum_{n} e^{-i p d n} a_{n}.
\end{equation}
Then the Hamiltonian $H$ in the single-excitation subspace $(\mathcal{N}=1)$ is rewritten as
\begin{equation}
H=\sum_{q} \Omega a_{q}^{\dagger} a_{q}+\sum_{k} \varepsilon_{k} \sigma_{k}^{\dagger} \sigma_{k}+N^{-1 / 2} \sum_{k, q} g_{k, q} \sigma_{k+q}^{\dagger} \sigma_{k}\left(a_{q}+a_{-q}^{\dagger}\right)+N^{-1} \sum_{k . q, q^{\prime}} g_{k, q, q^{\prime}}^{(2)} \sigma_{k+q+q^{\prime}}^{\dagger} \sigma_{k}\left(a_{q}+a_{-q}^{\dagger}\right)\left(a_{q^{\prime}}+a_{-q^{\prime}}^{\dagger}\right),
\end{equation}
where the polariton energy $\varepsilon_{k}$ and couplings $g_{k, q}, g_{k, q, q^{\prime}}^{(2)}$ are given by
\begin{equation}
\begin{gathered}
\varepsilon_{k}=\frac{\Gamma_{0}}{4} f(k)-i \frac{N \Gamma_{0}}{4} \delta_{k, \pm k_{0}}, \\
g_{k, q}=\frac{i}{4} \eta \Gamma_{0}\left(f_{-}(k)-f_{-}(k+q)\right)+\frac{1}{4} \eta \Gamma_{0} N\left(\delta_{k, k_{0}}-\delta_{k,-k_{0}}-\delta_{k, k_{0}-q}+\delta_{k,-k_{0}-q}\right), \\
g_{k, q, q^{\prime}}^{(2)}=-\frac{\Gamma_{0}}{8} \eta^{2}\left[f(k)+f\left(k+q+q^{\prime}\right)-f\left(k+q^{\prime}\right)-f(k+q)\right], \\
+i \frac{\Gamma_{0}}{8} \eta^{2} N \sum_{\pm}\left(\delta_{k, \pm k_{0}}+\delta_{k, \pm k_{0}-q-q^{\prime}}-2 \delta_{k, \pm k_{0}-q^{\prime}}\right),
\end{gathered}
\label{parameters}
\end{equation}
and
\begin{equation}
f(k)=\cot \frac{1}{2}\left(k_{0} d-k d\right)+\cot \frac{1}{2}\left(k_{0} d+k d\right), \quad f_{-}(k)=\cot \frac{1}{2}\left(k d-k_{0} d\right)+\cot \frac{1}{2}\left(k d+k_{0} d\right).
\end{equation}
Here the divergences encountered in $f$ and $f_{-}$can be simply neglected since their contributions have already been included in the Kronecker delta functions. We then resort to the Green's function technique in order to analyze the polariton-phonon hybrid excitations. Specifically, we focus on the subradiant excitations with $|k| \sim \pi / d$, i.e., near the edge of the Brillouin zone. The bare Green's functions of the polaritons and phonons are given respectively by
\begin{equation}
G^{(0)}(k, E)=\frac{1}{E-\varepsilon_{k}+i 0^{+}}, \quad D^{(0)}(q, \omega)=\frac{2 \Omega}{\omega^{2}-\Omega^{2}+i 0^{+}}.
\end{equation}
Taking into account the phonon dressing processes with contributions at $O\left(\eta^{2}\right)$, the polariton self-energy is given by $\Sigma=\Sigma_{d}+\Sigma_{s r}+\Sigma_{2}$, where
$$
\Sigma_{d(s r)}(k, E)=\frac{i}{2 \pi N} \int d \omega \sum_{q \neq(=) k \pm k_{0}} g_{k-q, q} g_{k,-q} G^{(0)}(k-q,-\omega+E) D^{(0)}(q, \omega)
$$
\begin{equation}
=\frac{1}{N} \sum_{q \neq(=) k \pm k_{0}} g_{k-q, q} g_{k,-q} \frac{1}{E-\Omega-\varepsilon_{k-q}+i 0^{+}},
\end{equation}
\begin{equation}
\Sigma_{2}(k, E)=\frac{i}{2 \pi N} \int d \omega \sum_{q} g_{k, q,-q}^{(2)} D^{(0)}(q, \omega)=\frac{1}{N} \sum_{q} g_{k, q,-q}^{(2)}.
\end{equation}
The self-energy contributions $\Sigma_{2}$ and $\Sigma_{s r}$ can be directly evaluated as
\begin{equation}
\begin{gathered}
\Sigma_{2}(k, E)=-i \frac{\Gamma_{0}}{2} \eta^{2}-\eta^{2} \varepsilon_{k,} \\
\Sigma_{s r}(k, E)=-\frac{1}{N} \sum_{\pm} g_{\mp k_{0}, k \pm k_{0}} g_{k,-k \mp k_{0}} \frac{1}{E-\Omega-\varepsilon_{\mp k_{0}}+i 0^{+}}=i \frac{1}{2} \eta^{2} \Gamma_{0}, \quad k \neq \pm k_{0}.
\end{gathered}
\end{equation}
During the calculation of $\Sigma_{s r}$, we adopt the approximation $E-\Omega-\varepsilon_{\mp k_{0}}+i 0^{+} \approx-\varepsilon_{\mp k_{0}}$ since $\varepsilon_{\mp k_{0}} \sim O(N)$ is much larger than other energy scales in the $N \rightarrow \infty$ limit. While evaluating the summations in $\Sigma_{d}$ is rather difficult, analytical results can be obtained in the $\varphi \ll 1$ limit. We simplify $k d$ in the expressions as $k$ and map the summations to the integrations from $-\pi$ to $\pi$. The imaginary parts of $\Sigma_{d}$ are given by
\begin{equation}
\operatorname{Im} \Sigma_{d}(k, E)=-\frac{1}{32} \eta^{2} \Gamma^{2} \int d q\left(f_{-}(k)-f_{-}(k-q)\right)^{2} \delta\left(E-\Omega-\varepsilon_{k-q}\right).
\end{equation}
The equation $E-\Omega-\varepsilon_{k-q}=0$ can be simplify as $\Omega+\varepsilon_{k-q}=0$ since the bare polariton energy $\varepsilon_{k} \ll \Omega$ when $|k| \sim \pi$ and $\Omega \sim \Gamma_{0}$, and $\varepsilon_{k}$ behaves as $\varepsilon_{k} \approx-\varphi \Gamma_{0} / k^{2}$ in the regime $\varphi \ll k<1$. Thus, we arrive at the solution $k-q=\left(\varphi \Gamma_{0} / \Omega\right)^{1 / 2}$ which indeed satisfies $\varphi \ll k-q \ll 1$. Moreover, in the same regime we have $\left|f_{-}(k)\right| \ll\left|f_{-}(k-q)\right|$ and $f_{-}(k-q) \approx 4 /(k-q)$. Thus we have
\begin{equation}
\operatorname{Im} \Sigma_{d}(k, E)=-\frac{1}{32} \eta^{2} \Gamma_{0}^{2} \int d q f_{-}^{2}(k-q) \frac{\delta\left(k-q-\left(\varphi \Gamma_{0} / \Omega\right)^{1 / 2}\right)}{\left|\partial \varepsilon_{k^{\prime}} /\left.\partial k^{\prime}\right|_{k^{\prime}=k-q}\right|}=-\frac{1}{4} \eta^{2} \Gamma_{0}\left(\frac{\Gamma_{0}}{\Omega \varphi}\right)^{1 / 2}.
\end{equation}
The real parts of $\Sigma_{d}$ read
\begin{equation}
\begin{aligned}
\label{real_self_energy_integral}
\operatorname{Re} \Sigma_{d}(k, E)=\frac{1}{16} & \eta^{2} \Gamma_{0}^{2} \mathcal{P} \int \frac{d q}{2 \pi} \frac{\left(f_{-}(k)-f_{-}(k-q)\right)^{2}}{E-\Omega-\varepsilon_{k-q}} \\
&=\frac{1}{16} \eta^{2} \Gamma_{0}^{2}\left[f_{-}^{2}(k) \mathcal{P} \int \frac{d q}{2 \pi} \frac{1}{E-\Omega-\varepsilon_{q}}-\frac{1}{\Omega-E} \mathcal{P} \int \frac{d q}{2 \pi} \frac{f_{-}^{2}(q)}{1+(\Omega-E)^{-1} \varepsilon_{q}}\right],
\end{aligned}
\end{equation}
where $\mathcal{P}$ denotes the Cauchy principal value. The first integral can be evaluated approximately as $-\Omega^{-1}$ since $E-\Omega-\varepsilon_{q} \approx-\Omega$ except for $q \ll 1$, which only constitutes a narrow integral of $[-\pi, \pi]$. For the second integral, it can be verified that this integral behaves almost as a constant. Thus we have
\begin{equation}
\operatorname{Re} \Sigma_{d}(k, E) \approx-\frac{1}{16} \eta^{2} \Gamma_{0}^{2}\left[\frac{1}{\Omega}(\pi-|k|)^{2}+\frac{\alpha}{\Omega-E}\right], \quad|k| \sim \pi.
\end{equation}
where $\alpha$ is the value of the second integral in Eq. (\ref{real_self_energy_integral}) and can be evaluated numerically as $\alpha \approx 4$. These results allow us to obtain the $Z$ factor and effective mass as
\begin{equation}
Z^{-1} \approx 1+\frac{\alpha \Gamma_{0}^{2}}{16 \Omega^{2}} \eta^{2}, \quad \frac{m}{m^{*}} \approx 1+\left(\frac{\Gamma_{0}}{\varphi \Omega}-\frac{\alpha \Gamma_{0}^{2}}{16 \Omega^{2}}-1\right) \eta^{2},
\end{equation}
where the bare effective mass $m$ can be determined by the dispersion relation near the Brillouin zone edge $\varepsilon_{k}-\varepsilon_{\pi} \approx-\varphi \Gamma_{0}(k-\pi)^{2} / 16$. In the $\varphi \ll 1$ limit, the effective mass is approximated by $m^{*} \approx m /\left(1+\eta^{2} \Gamma_{0}(\varphi \Omega)^{-1}\right)$.

\section{Schrieffer-Wolff transformation in finite arrays}
\begin{figure}
            \centering
            \includegraphics[width=\columnwidth]{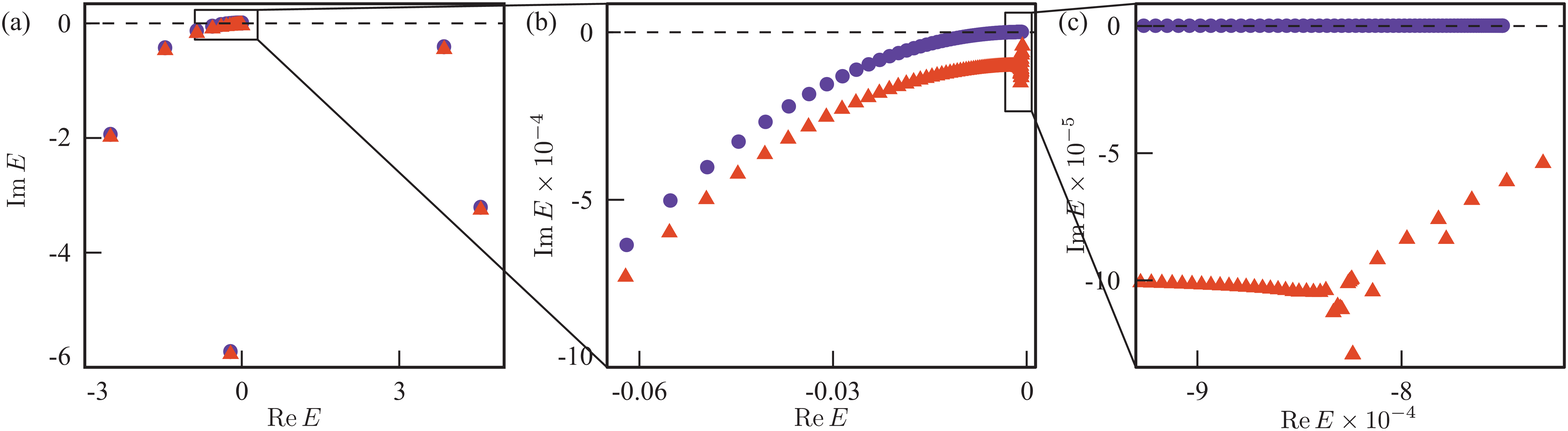}
            \caption{
               Complex single-excitation spectrums obtained from the Schrieffer-Wolff transformation results Eq. (\ref{effective_correction_0})
(red triangle) in comparison with the single-polariton spectrums (blue circle). The spectrums are zoomed in different scales. Here the system parameters are selected to be the same as Fig. 2(a) in the main text: $10\Gamma_{0}=\Omega=1$, $\varphi=0.03$, $\eta / \varphi=1$, and $N=240$.
            }
            \label{spectrum_a}
\end{figure}

\begin{figure}
            \centering
            \includegraphics[width=\columnwidth]{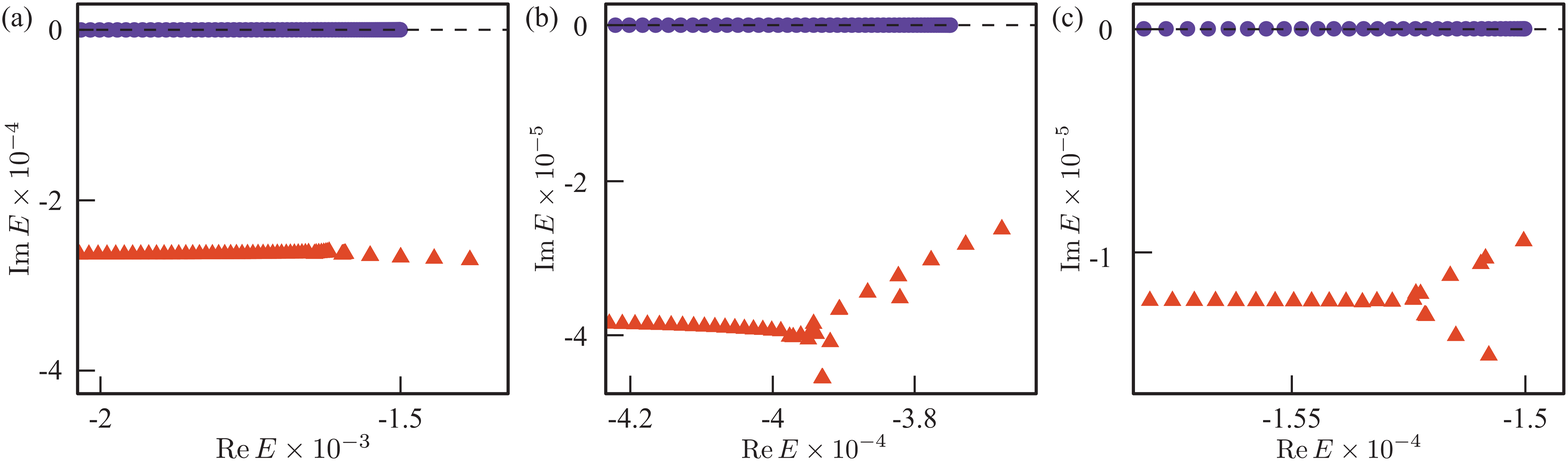}
            \caption{
               The most subradiant part of the complex single-excitation spectrum obtained from the Schrieffer-Wolff transformation results Eq. (\ref{effective_correction_0})
(red triangle) in comparison with the single-polariton spectrums (blue circle). The calculation has been performed for $\Omega=1$, $\eta / \varphi=1$, $N=240$ and $\Gamma_{0}=0.1$, $\varphi=0.06$ in (a), $\Gamma_{0}=0.05$, $\varphi=0.03$ in (b), and $\Gamma_{0}=0.02$, $\varphi=0.03$ in (c) respectively.
            }
            \label{spectrum_b}
\end{figure}

In this section, we present a detailed derivation for the Schrieffer-Wolff transformation (SWT) performed in finite arrays to decoupled the polaritons from phonons. This is done by finding a proper operator $S$ to cancel the couplings in $e^{S} H e^{-S}$ to the first order, then projecting the transformed Hamiltonian to the single-excitation subspace without phonons \cite{BRAVYI20112793,coleman_2015}. The effective Hamiltonian describing polaritons and phonons are given by $H=H_{0}+H_{p}+H_{I 1}+H_{I 2}$, where
\begin{equation}
\begin{gathered}
H_{0}=-i \frac{\Gamma_{0}}{2} \sum_{m, n} e^{i \varphi|m-n|} \sigma_{m}^{\dagger} \sigma_{n}, \quad H_{p}=\sum_{m} \Omega a_{m}^{\dagger} a_{m}, \\
H_{I 1}=\frac{1}{2} \eta \Gamma_{0} \sum_{m, n} \operatorname{sign}(m-n) e^{i k_{0}\left|z_{m}-z_{n}\right|} \sigma_{m}^{\dagger} \sigma_{n}\left(a_{m}+a_{m}^{\dagger}-a_{n}-a_{n}^{\dagger}\right), \\
H_{I 2}=-i \frac{1}{4} \eta^{2} \Gamma \sum_{m, n}^{N} e^{i k_{0}\left|z_{m}-z_{n}\right|} \sigma_{m}^{\dagger} \sigma_{n}\left(a_{m}+a_{m}^{\dagger}-a_{n}-a_{n}^{\dagger}\right)^{2}.
\end{gathered}
\end{equation}
Here we consider the emitter-phonon coupling up to the second order in the emitter displacements. The emitter Hamiltonian $H_{0}$ can be diagonalized as
\begin{equation}
H_{0}=\sum_{\tilde{k}} \varepsilon_{\tilde{k}}^{\prime}\left|\psi_{\tilde{k}}\right\rangle\left\langle\psi_{\tilde{k}}|, \quad| \psi_{\tilde{k}}\right\rangle \propto g_{-\tilde{k}}|\widetilde{k}, 0\rangle-g_{\tilde{k}}|-\tilde{k}, 0\rangle,
\end{equation}
where $\varepsilon_{\tilde{k}}^{\prime}=\epsilon_{\tilde{k}}-i \gamma_{\tilde{k}}$ is the eigenvalue of the state $\left|\psi_{\tilde{k}}\right\rangle$, the wavenumber $\tilde{k}$ satisfies the equation $g_{\tilde{k}} h_{-\tilde{k}}=g_{-\tilde{k}} h_{\tilde{k}}$, $g_{\tilde{k}}=e^{i\left(\tilde{k}-k_{0}\right) z_{1}} /\left[1-e^{i\left(\tilde{k}-k_{0}\right) d}\right]$ and $h_{k}=e^{i\left(k+k_{0}\right) z_{N}} /\left[e^{-i\left(k+k_{0}\right) d}-1\right]$ \cite{PhysRevLett122203605,PhysRevLett125253601}. The operator $S$ should satisfy the condition $\left[H_{0}+H_{p}, S\right]=H_{I 1}$ in order to remove the couplings to the first order, and is chosen as
\begin{equation}
S=N^{-1 / 2} \sum_{\tilde{k}_{1}, \tilde{k}_{2}, q}\left|\psi_{\tilde{k}_{1}}\right\rangle\left\langle\psi_{\tilde{k}_{2}}\right|\left(A_{\tilde{k}_{1}, \tilde{k}_{2}, q} a_{q}+B_{\tilde{k}_{1}, \tilde{k}_{2}, q} a_{-q}^{\dagger}\right).
\end{equation}
The couplings $H_{I 1}$ can be written as $H_{I 1}=N^{-1 / 2} \sum_{\tilde{k}_{1}, \tilde{k}_{2}, q} \tilde{g}_{\tilde{k}_{1}, \tilde{k}_{2}, q}\left|\psi_{\tilde{k}_{1}}\right\rangle\left\langle\psi_{\tilde{k}_{2}}\right|\left(a_{q}+a_{-q}^{\dagger}\right)$, thus the coefficients $A$ and $B$ can be obtained as
\begin{equation}
A_{\tilde{k}_{1}, \tilde{k}_{2}, q}=\frac{\tilde{g}_{\tilde{k}_{1}, \tilde{k}_{2}, q}}{-\Omega+\varepsilon_{\tilde{k}_{1}}^{\prime}-\varepsilon_{\tilde{k}_{2}}^{\prime}}, \quad B_{\tilde{k}_{1}, \tilde{k}_{2}, q}=\frac{\tilde{g}_{\tilde{k}_{1}, \tilde{k}_{2}, q}}{\Omega+\varepsilon_{\tilde{k}_{1}}^{\prime}-\varepsilon_{\tilde{k}_{2}}^{\prime}}.
\end{equation}
Therefore, the effective Hamiltonian to the second order of $\eta$ is given by
\begin{equation}
\label{effective_correction_0}
\begin{gathered}
H^{\prime}=H_{0}+\frac{1}{2} P_{0}\left(\left[S, H_{I 1}\right]+H_{I 2}\right) P_{0}=\left(1-\eta^{2}\right) H_{0}-i \frac{1}{2} \eta^{2} \Gamma_{0}+\sum_{\tilde{k}_{1}, \tilde{k}_{2}} \Delta_{\tilde{k}_{1}, \tilde{k}_{2}}\left|\psi_{\tilde{k}_{1}}\right\rangle\left\langle\psi_{\tilde{k}_{2}}\right|, \\
\Delta_{\tilde{k}_{1}, \tilde{k}_{2}}=\frac{1}{2 N} \sum_{q, \tilde{k}} \tilde{g}_{\tilde{k}, \tilde{k}_{2},-q}\left(A_{\tilde{k}_{1}, \tilde{k}, q}-B_{\tilde{k}_{1}, \tilde{k}_{,} q}\right),
\end{gathered}
\end{equation}
where $P_{0}$ is the projector onto the subspace with zero phonons. 
The excitation spectrum can be obtained by diagonalizing $H^{\prime}$, as shown in Figs. \ref{spectrum_a} and \ref{spectrum_b}. 
It is clear that the spectrum is modified significantly in the most subradiant regime compared to the polariton spectrum.

To complete the derivation, we need to ensure the validity of the Schrieffer-Wolff transformation. In order to perform the Schrieffer-Wolff transformation correctly, the single-polariton subspace without phonons should be separated from the rest of the spectrum by an energy gap, such that the strength of the interaction Hamiltonian $H_{I 1}$ is much smaller than the energy gap \cite{BRAVYI20112793,coleman_2015}. This condition can be written as
\begin{equation}
|\langle\psi_{\tilde{k}^{\prime}}, q|H_{I 1}| \psi_{\tilde{k}}\rangle| \ll|\varepsilon_{\tilde{k}^{\prime}, q}^{\prime}-\varepsilon_{\tilde{k}}^{\prime}|,
\label{condition0}
\end{equation}
where $\varepsilon_{\tilde{k}}^{\prime}$ and $\varepsilon_{\tilde{k}^{\prime},q}^{\prime}=\varepsilon_{\tilde{k}^{\prime}}^{\prime}+\Omega$ are the eigenvalues of the states $|\psi_{\tilde{k}}\rangle$ and $|\psi_{\tilde{k}^{\prime}}, q\rangle=a_q^{\dagger}|\psi_{\tilde{k}^{\prime}}\rangle$ respectively.

To proceed, we note that the eigenvalue $\varepsilon_{\tilde{k}}^{\prime}$ can be approximated by
\begin{equation}
\varepsilon_{\tilde{k}}^{\prime}=\epsilon_{\tilde{k}}-i \gamma_{\tilde{k}} \approx \varepsilon_k-i \gamma_{\tilde{k}},
\end{equation}
where $\varepsilon_k$ is the eigenvalue of $H_0$ in an infinite array and $k=\operatorname{Re} \tilde{k}$. 
Meanwhile, the matrix element of $H_{I 1}$ can be estimated by
\begin{equation}
|\langle\psi_{\tilde{k}^{\prime}},q|H_{I 1}| \psi_{\tilde{k}}\rangle| \sim|\langle k^{\prime}, q|H_{I 1}| k, 0\rangle| \approx N^{-1 / 2} g_{k,-q} \delta_{k^{\prime}, k-q},
\end{equation}
since $H_{I 1}$ perturbs the single-polariton state $|k, 0\rangle=\sigma_k^{\dagger}|0\rangle$ and only connects it to $|k-q, q\rangle=\sigma_{k-q}^{\dagger} a_q^{\dagger}|0\rangle$ in an infinite array.

In infinite arrays, the condition Eq. (\ref{condition0}) fails near $q=k \pm k_{\Omega}$ where $k_{\Omega}>0$ is determined by $\varepsilon_k=$ $\Omega+\varepsilon_{k_{\Omega}}$, since the single-polariton state $|k, 0\rangle$ is degenerated with $\left|\pm k_{\Omega}, k \mp k_{\Omega}\right\rangle$. However, this degeneracy is lifted in finite arrays due to the discrepancy in their decay rates. There exists an imaginary energy gap $\Delta \approx i(\gamma_{\tilde{k}_{\Omega}}-\gamma_{\tilde{k}})$ separating states $|\psi_{\tilde{k}}\rangle$ and $|\psi_{\pm \tilde{k}_{\Omega}}, k \mp k_{\Omega}\rangle$. Therefore, the condition Eq. (\ref{condition0}) can be satisfied if this energy gap is much larger than the corresponding matrix element of $H_{I 1}$. This is equivalent to
\begin{equation}
|\gamma_{\tilde{k}_{\Omega}}-\gamma_{\tilde{k}}| \gg|\langle\pm k_{\Omega}, k \mp k_{\Omega}|H_{I 1}| k, 0\rangle|.
\end{equation}
For subradiant excitation $k \approx \pi / d$, its decay rate scales as \cite{PhysRevLett122203605}
\begin{equation}
\gamma_{\tilde{k}} \sim N^{-1} \Gamma_0 \varphi^2(k-\pi / d)^2.
\end{equation}
Meanwhile, the decay rate $\gamma_{\tilde{k}_{\Omega}}$ can be calculated as follows. Supposing $\tilde{k}_{\Omega}=k_{\Omega}+\delta / N$, then $\delta$ describes the correction to the wavenumber. According to the equation $g_{\tilde{k}_{\Omega}} h_{-\tilde{k}_{\Omega}}=g_{-\tilde{k}_{\Omega}} h_{\tilde{k}_{\Omega}}$,
we can find that $\delta \approx-i \varphi /\left(k_{\Omega} d\right)$ when $\varphi \ll k_{\Omega} d \ll 1$. Next, we substitute $\tilde{k}_{\Omega}$ into the expression for $\varepsilon_k$ (Eq. (\ref{parameters})), then this imaginary correction leads to the decay rate
\begin{equation}
\gamma_{\tilde{k}_{\Omega}} \sim N^{-1} \Omega^2 \Gamma_0^{-1} .
\end{equation}
It is clear that $\gamma_{\tilde{k}_{\Omega}} \gg \gamma_{\tilde{k}}$, thus the spectral gap $\Delta$ scales as $\Delta \approx i \gamma_{\tilde{k}_{\Omega}} \sim i N^{-1} \Omega^2 \Gamma_0^{-1}$. Meanwhile, we have
\begin{equation}
\left|\left\langle\pm k_{\Omega}, k \mp k_{\Omega}\left|H_{I 1}\right| k, 0\right\rangle\right|=N^{-1 / 2} g_{k, \pm k_{\Omega}-k} \approx N^{-1 / 2} \eta \Gamma_0\left(\varphi \Gamma_0 / \Omega\right)^{-1 / 2},
\end{equation}
in the same regime $\varphi \ll k_{\Omega} d \ll 1$. Therefore, the condition Eq. (\ref{condition0}) is equivalent to
\begin{equation}
\eta^2 \varphi^{-1}\left(\Gamma_0 / \Omega\right)^3 \ll N^{-1} .
\label{condition}
\end{equation}

Hence, the Schrieffer-Wolff transformation is well-defined and provide correct results in the weak coupling regime $\eta \ll 1$ if $\eta \varphi^{-1}\left(\Gamma_0 / \Omega\right)^3$ scales as $N^{-s}(s \geq 1)$. This is achievable in experiments since both the decay rate $\Gamma_0$ and the atomic spacing $d$ are highly tunable. For example, we can maintain $\Gamma_0 / \Omega \sim N^{-1 / 3}$ while $\eta$ and $\varphi$ are remained unchanged as the system size $N$ increases. In all figures of the main text, the parameters are chosen such that the condition Eq. (\ref{condition}) is fulfilled. This justifies the validity of the Schrieffer-Wolff transformation in our work.


\section{The emergence of quasiperiodic structure}
\begin{figure}
            \centering
            \includegraphics[scale=0.5]{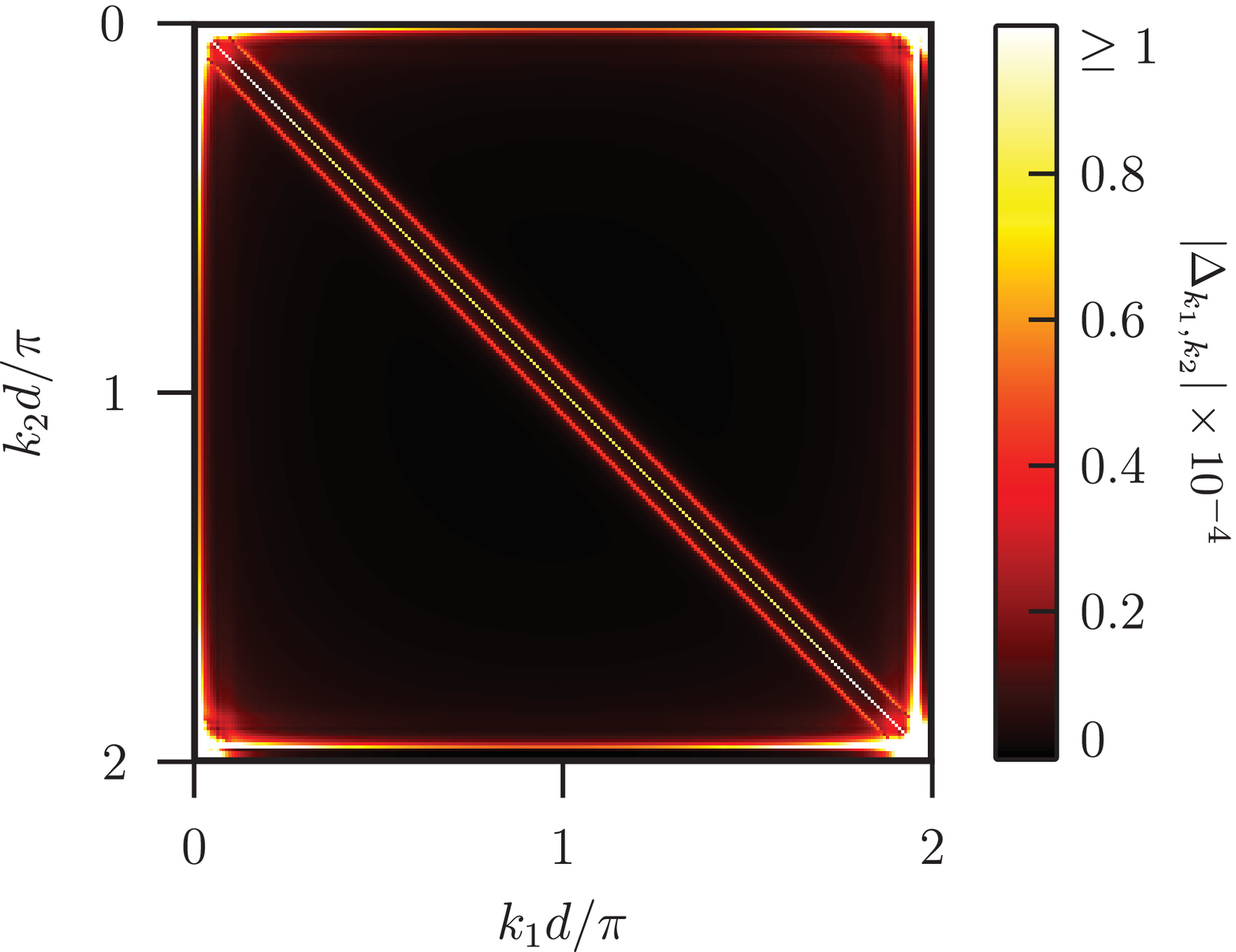}
            \caption{
               Color map of the phonon correction $\Delta$ in $k$-space. The calculation has been performed for $10\Gamma_{0}=\Omega=1$, $\varphi=0.06$, $\eta / \varphi=1$, and $N=240$.
            }
            \label{Delta_k}
\end{figure}

In this section, we focus on the effects of resonant phonon processes on the phonon correction $\Delta=$ $\sum \Delta_{\tilde{k}_{1}, \tilde{k}_{2}}\left|\psi_{\tilde{k}_{1}}\right\rangle\left\langle\psi_{\tilde{k}_{2}}\right|$, and present a derivation for the approximate effective Hamiltonian $H_{eff}$ in the main text. We first replace the discrete $\operatorname{sum} \sum_{\tilde{k}}$ in Eq. (\ref{effective_correction_0}) by the integral $(2 \pi)^{-1} \int d \tilde{k}$. In the expression
\begin{equation}
\frac{1}{-\Omega+\varepsilon_{\tilde{k}_{1}}^{\prime}-\varepsilon_{\tilde{k}_{2}}^{\prime}}=\frac{1}{-\Omega+\epsilon_{\tilde{k}_{1}}-\epsilon_{\tilde{k}_{2}}+i\left(\gamma_{\tilde{k}_{2}}-\gamma_{\tilde{k}_{1}}\right)},
\end{equation}
the decay rates $\gamma_{\tilde{k}_{1}}$ and $\gamma_{\tilde{k}_{2}}$ are much smaller than other energy scales and satisfy $\gamma_{\tilde{k}_{2}}>\gamma_{\tilde{k}_{1}}$ for subradiant states on the lower branch of the spectrum with $-\Omega+\epsilon_{\tilde{k}_{1}}-\epsilon_{\tilde{k}_{2}}=0$. Therefore, we can replace $i\left(\gamma_{\tilde{k}_{2}}-\gamma_{\tilde{k}_{1}}\right)$ by $i 0^{+}$and apply the Sokhotski formula
\begin{equation}
\frac{1}{-\Omega+\varepsilon_{\tilde{k}_{1}}^{\prime}-\varepsilon_{\tilde{k}_{2}}^{\prime}} \approx \mathcal{P} \frac{1}{-\Omega+\epsilon_{\tilde{k}_{1}}-\epsilon_{\tilde{k}_{2}}}-i \pi \delta\left(-\Omega+\epsilon_{\tilde{k}_{1}}-\epsilon_{\tilde{k}_{2}}\right).
\end{equation}
Thus, we can separate the phonon correction as $\Delta=\Delta^{\prime}+\Delta^{\prime \prime}$, where
\begin{equation}
\Delta_{\tilde{k}_{1}, \tilde{k}_{2}}^{\prime}=\frac{1}{4 \pi} \sum_{q} \mathcal{P} \int d \tilde{k} \tilde{g}_{\tilde{k}, \tilde{k}_{2},-q}\left(A_{\tilde{k}_{1}, \tilde{k}, q}-B_{\tilde{k}_{1}, \tilde{k}, q}\right),
\end{equation}
describes the energy shift to the excitation, and
\begin{equation}
\Delta_{\tilde{k}_{1}, \tilde{k}_{2}}^{\prime \prime}=-\frac{i}{4} \sum_{q} \int d \tilde{k} \tilde{g}_{\tilde{k}, \tilde{k}_{2},-q} \tilde{g}_{\tilde{k}_{1}, \tilde{k}, q}\left(\delta\left(-\Omega+\epsilon_{\tilde{k}_{1}}-\epsilon_{\tilde{k}}\right)+\delta\left(-\Omega+\epsilon_{\tilde{k}_{2}}-\epsilon_{\tilde{k}}\right)\right).
\end{equation}
Similar to the previous discussions, the solution of $-\Omega+\epsilon_{\tilde{k}_{1(2)}}-\epsilon_{\tilde{k}}=0$ can be approximately given by $\tilde{k}=k_{\Omega}$ if we consider phonon corrections to the states in  quasi-flat regime. In the $k$-space, the corrections $\Delta^{\prime \prime}$ can be evaluated as
\begin{equation}
\begin{gathered}
\Delta_{k_{1}, k_{2}}^{\prime \prime}=\sum_{\tilde{k}_{1}, \bar{k}_{2}} \Delta_{\tilde{k}_{1}, \tilde{k}_{2}}^{\prime \prime}\left\langle k_{1} \mid \psi_{\tilde{k}_{1}}\right\rangle\left\langle\psi_{\tilde{k}_{2}} \mid k_{2}\right\rangle=-\frac{i}{2\left|\varepsilon_{k_{\Omega}}^{\prime}\right|} \sum_{\tilde{k}_{1}, \bar{k}_{2}, q} \tilde{g}_{k_{\Omega}, \bar{k}_{2,-q}} \tilde{g}_{\tilde{k}_{1}, k_{\Omega}, q}\left\langle k_{1} \mid \psi_{\tilde{k}_{1}}\right\rangle\left\langle\psi_{\tilde{k}_{2}} \mid k_{2}\right\rangle \\
=-\frac{i}{2\left|\varepsilon_{k_{\Omega}}^{\prime}\right|} \sum_{q, k, k^{\prime}, \tilde{k}_{1}, \bar{k}_{2}} g_{k,-q} g_{k^{\prime}, q}\left\langle\psi_{k_{\Omega}} \mid k-q\right\rangle\left\langle k \mid \psi_{\tilde{k}_{2}}\right\rangle\left\langle\psi_{\tilde{k}_{2}} \mid k_{2}\right\rangle\left\langle k_{1} \mid \psi_{\tilde{k}_{1}}\right\rangle\left\langle\psi_{\tilde{k}_{1}} \mid k^{\prime}+q\right\rangle\left\langle k^{\prime} \mid \psi_{k_{\Omega}}\right\rangle \\
=-\frac{i}{2\left|\varepsilon_{k_{\Omega}}^{\prime}\right|} \sum_{q} g_{k_{2},-q} g_{k_{1}-q, q}\left\langle\psi_{k_{\Omega}} \mid k_{2}-q\right\rangle\left\langle k_{1}-q \mid \psi_{k_{\Omega}}\right\rangle.
\end{gathered}
\end{equation}
The eigenstate $\left|\psi_{k_{\Omega}}\right\rangle$ can be approximated by $\left|\psi_{k_{\Omega}}\right\rangle \approx\left(\left|k_{\Omega}\right\rangle+\left|-k_{\Omega}\right\rangle\right) / \sqrt{2}$ since $g_{\tilde{k}} \approx-g_{-\tilde{k}}$ when $\varphi \ll$ $k_{\Omega} \ll 1$. Thus,
\begin{equation}
\Delta_{k_{1}, k_{2}}^{\prime \prime}=-\frac{i}{8} \eta^{2} \Gamma_{0}\left(\frac{\Gamma_{0}}{\Omega \varphi}\right)^{\frac{1}{2}}\left(2 \delta_{k_{1}, k_{2}}+\delta_{k_{1}, k_{2}+2 k_{\Omega}}+\delta_{k_{1}, k_{2}-2 k_{\Omega}}\right),
\end{equation}
where we use the facts that $g_{k_{\Omega}-q, q} \approx-g_{k_{\Omega, q}}=\mathrm{i} \eta \Gamma_{0} f\left(k_{\Omega}\right) / 4$. Thus, we have
\begin{equation}
\label{ideal_correction}
\Delta^{\prime \prime}=-\frac{i}{8} \eta^{2} \Gamma_{0}\left(\frac{\Gamma_{0}}{\Omega \varphi}\right)^{\frac{1}{2}} \sum_{k}\left(2|k\rangle\langle k|+| k\rangle\left\langle k+2 k_{\Omega}|+| k+2 k_{\Omega}\right\rangle\langle k|\right).
\end{equation}
For a finite array, the Sokhotski formula cannot exactly describe the behavior of $\left(-\Omega+\varepsilon_{\tilde{k}_{1}}^{\prime}-\varepsilon_{\tilde{k}_{2}}^{\prime}\right)^{-1}$ due to the discreteness of energy spectrum. This leads to the deviations from the Eq. (\ref{ideal_correction}), and requires the introductions of $k$-dependent coefficients. As a result, the phonon correction $\Delta$ has the form
\begin{equation}
\label{effective_correction}
\Delta \approx \sum_{k}\left[V_{k}\left(e^{-i \theta_{k}}|k\rangle\langle k+2 k_{\Omega}|+e^{i \theta_{k}}| k+2 k_{\Omega}\rangle\langle k|\right) / 2+\delta \varepsilon_{k}|k\rangle\langle k|\right].
\end{equation}
where $V_{k}$ and $\theta_{k}$ are $k$-dependent coupling strength and phase factor respectively, and $\delta \varepsilon_{k}$ is the energy shift. We can verify that the phonon correction are well approximated by Eq. (\ref{effective_correction}), as shown in Fig. \ref{Delta_k}. It is clear that the main contributions come from $|k\rangle\langle k|$ and $|k\rangle\langle k \pm 2 k_{\Omega}|$, which manifest themself as the oblique lines parallel to the diagonal in the figure. 

Coefficients $V_{k}$, $\theta_{k}$ and $\delta \varepsilon_{k}$ change slowly in the subradiant regime, thus we can replace them by their values at $k=\pi / d$ if we only concern about subradiant excitations. 
Omitting the energy shift $\delta \varepsilon = \delta \varepsilon_{k=\pi / d}$, the resulting effective Hamiltonian in real space can be written as
\begin{equation}
H^{\prime} \approx H_{e f f}=H_{0}+V \sum_{m} \cos (2 \pi \beta m+\theta)| m\rangle\langle m|,
\label{effective_Hamiltonian_1}
\end{equation}
where $V=V_{k=\pi / d}$, $\theta=\theta_{k=\pi / d}$, and $\beta=k_{\Omega} d / \pi$ serves as the frequency. 

\begin{figure}
            \centering
            \includegraphics[width=\columnwidth]{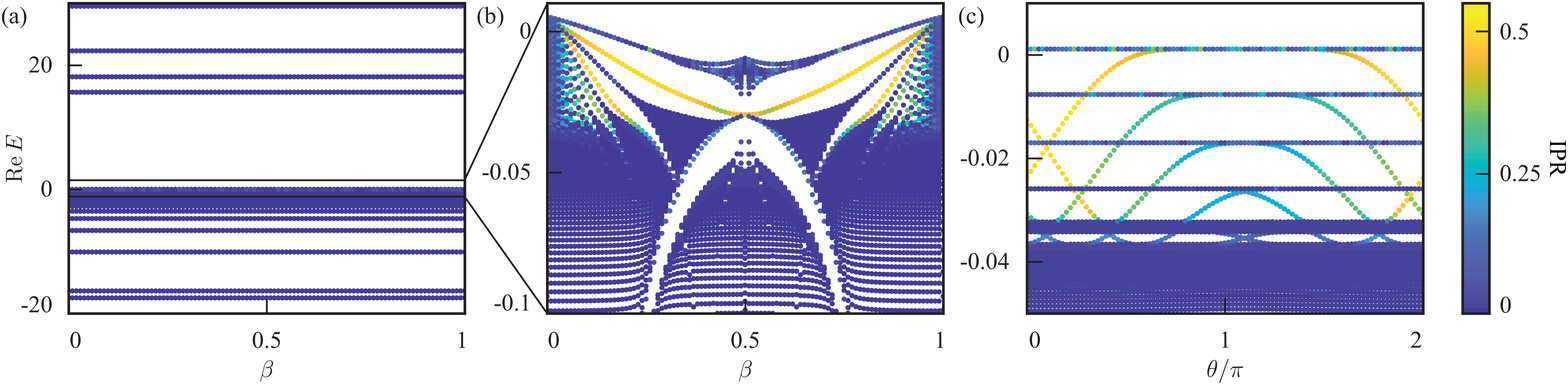}
            \caption{
               (a,b) Real spectrum of $H_{\text {eff }}$ (Eq. (\ref{effective_Hamiltonian_1})) zoomed in different scales as a function of $\beta$
 with $\theta=0$. (c) Real spectrum of the most subradiant excitations in $H_{\text {eff }}$ as a function of  $\theta$ with $\beta=1 /(10+\beta)=\sqrt{26}-5$. 
All of these figures were generated for an array of $N=200$ emitters with $\Gamma_{0}=1$, $\varphi=0.059$ and $V=0.02$.
            }
            \label{butterfly}
\end{figure}

$H_{eff}$ can be regarded as a one-dimensional quasiperiodic model since $\beta$ is determined by the transcendental equation $\varepsilon_{k}=\Omega+\varepsilon_{k_{\Omega}}$ and is in general irrational. 
Due to the quasiperiodicity, the spectrum of $H_{eff}$ becomes very rich and exhibits a characteristic Hofstadter butterfly pattern, as shown in Fig. \ref{butterfly}(b). Furthermore, the spectrum also includes highly localized edge states which cross the spectral gaps by changing the modulation phase $\theta$ (Fig. \ref{butterfly}(c)). These properites clearly hint the topological nature of the spectral gaps and edge states.


\section{Multifractality in the effective Hamiltonian $H_{eff}$}
In this section, we perform a multifractal analysis on the effective Hamiltonian $H_{e f f}$ and reveal that it presents a similar ergodic-multifractal transition to the system we considered. 
The effective Hamiltonian $H_{e f f}$ reads
\begin{equation}
H_{e f f}=-i \frac{\Gamma_{0}}{2} \sum_{m, n} e^{i \varphi|m-n|}|m\rangle\langle n|+V \sum_{m} \cos (2 \pi \beta m+\theta)| m\rangle\langle m|,
\label{effective_Hamiltonian_2}
\end{equation}
where $\beta$ is the modulation frequency. 

Similar to the main text, we employ the fractal dimensions and the level spacings to distinguish ergodic and multifractal regions of the spectrum. 
We first compute the even-odd (odd-even) energy spacing $S_{n}^{e-o}=E_{2 n}-E_{2 n-1}$ ($S_{n}^{o-e}=E_{2 n+1}-E_{2 n}$) in Figs. \ref{multifractal}(a) and \ref{multifractal}(b), where $E_{n}$ are the real parts of eigenenergies sorted in ascending order. 
It is clear that all states are ergodic when the potential strength $V$ is small. 
For larger $V$, we can find that several bands become multifractal and exhibit strongly scattered distributions. 
An ergodic-to-multifractal edge separating two regions can also be identified. 

To determine the fractal dimension, we follow the standard box-counting procedure by dividing the system into $N / l$ boxes of size $l$ \cite{PhysRevB68024206,RevModPhys93045001,PhysRevB104224204}. 
For a normalized wavefunction $|\psi\rangle=\sum_{n} \psi_{n}|n\rangle$, the probability in the $i$-th box is given by $\mu_{i}=\sum_{n}\left|\psi_{n}\right|^{2}$ where the summation is performed inside the $i$-th box. 
The spectrum of fractal dimensions is given by
\begin{equation}
f\left(\alpha_{q}\right)=\lim _{\delta \rightarrow 0} \frac{\sum_{i=1}^{N / l} \mu_{i}(q) \ln \mu_{i}(q)}{\ln \delta}, \quad \alpha_{q}=\lim _{\delta \rightarrow 0} \frac{\sum_{i=1}^{N / l} \mu_{i}(q) \ln \mu_{i}(1)}{\ln \delta},
\end{equation}
where $\delta=l / N, \mu_{i}(q)=\mu_{i}^{q} / \sum_{i=1}^{N / l} \mu_{i}^{q}$ is the normalized $q$-th moment.
Meanwhile, $|\psi\rangle$ can be characterized by the moment $I_{q}=\sum_{n}\left|\psi_{n}\right|^{2 q} \propto N^{-\tau(q)}$, where $\tau(q)$ is related to $f\left(\alpha_{q}\right)$ via the Legendre transform $\tau(q)=q \alpha_{q}-f\left(\alpha_{q}\right)$. 
The fractal dimension $D_{q}$ is given by $D_{q}=\tau(q) /(q-1)$. 
In Fig. \ref{multifractal}(c), we show the mean fractal dimensions $\bar{D}_{2}$ for three bands with highest energy. 
It can be found that the ergodic-multifractal transitions happened at different potential strength $V$ for each band, and bands with lower energies require larger $V$ to become multifractal. 
These analyses confirm the presence of ergodic-multifractal transitions in the effective Hamiltonian $H_{e f f}$. 
\begin{figure}
            \centering
            \includegraphics[width=\columnwidth]{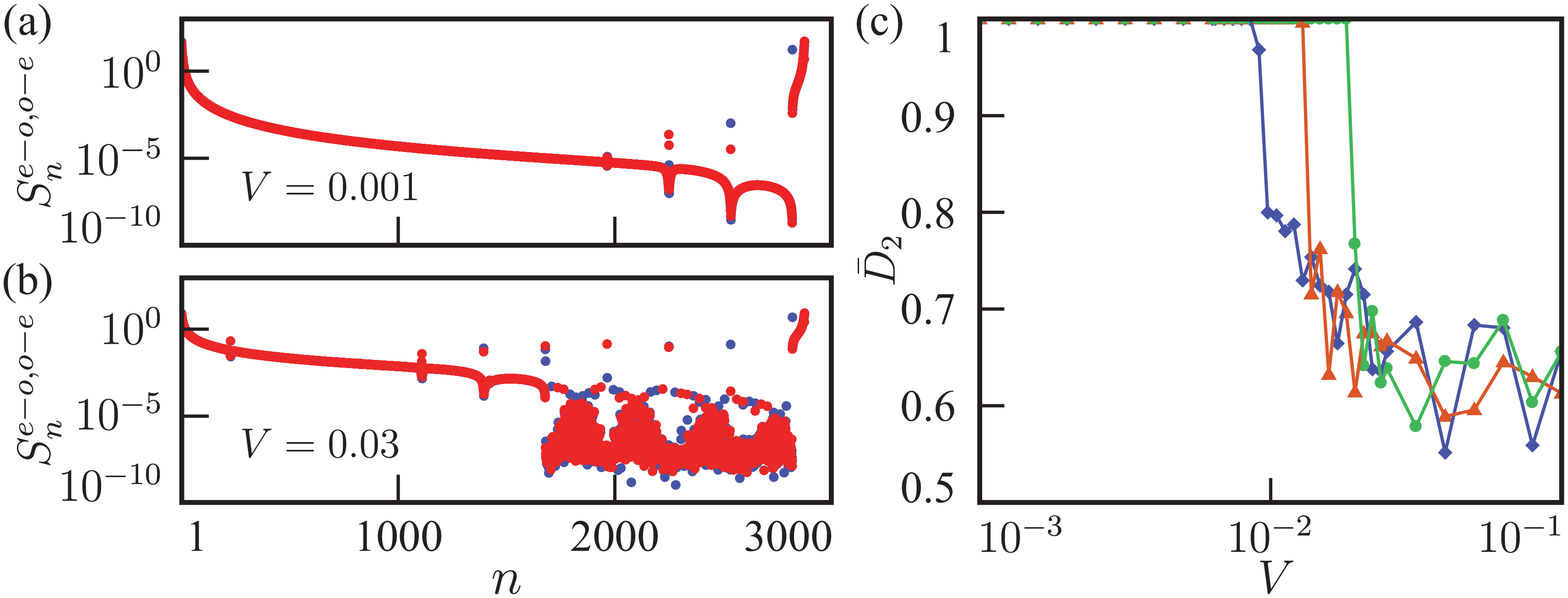}
            \caption{
               (a, b) Level spacing $S_{n}^{e-o}$ (red) and $S_{n}^{o-e}$ (blue) for the effective Hamiltonian $H_{e f f}$ (Eq. (\ref{effective_Hamiltonian_2})) with $V=0.001$ (a) and $V=0.03$ (b) respectively. (c). Mean fractal dimensions $\bar{D}_{2}$ for three subbands with highest energy. Three subbands (blue diamond, red triangle, green circle) are arranged in energy-descending order. All of these figures were generated for an array of $N=5760$ emitters with $\Gamma_{0}=\Omega=1, \varphi=\pi / 50, \beta=1 /(10+\beta)=\sqrt{26}-5$ and $\theta=0$.
            }
            \label{multifractal}
\end{figure}

\end{document}